\documentclass[twocolumn,showpacs,preprintnumbers,amsmath,amssymb]{revtex4-1}
\usepackage{dcolumn}
\usepackage{bm}
\usepackage[dvips]{graphicx}
\usepackage{wrapfig,subfigure}
\usepackage{color}

\newcommand{\cn}{{\rm cn}}

\newcommand{\dn}{{\rm dn}}

\date{\today}

\begin{document}
\title{Sharp tunneling peaks in a parametric oscillator: quantum resonances missing in the rotating wave approximation}
\author{V. Peano$^{(1)}$, M. Marthaler$^{(2)}$, and M. I. Dykman$^{(1)}$}
\affiliation{$^{(1)}$ Department of Physics and Astronomy, Michigan State University, East Lansing, Michigan 48824\\
$^{(2)}$ Institut f\"ur Theoretische Festk\"orperphysik
and DFG-Center for Functional Nanostructures (CFN), Karlsruhe Institute of Technology, D-76128 Karlsruhe, Germany}
\begin{abstract}
We describe a new mechanism of tunneling between period-two vibrational states of a weakly nonlinear parametrically modulated oscillator. The tunneling  results from resonant transitions induced by the fast oscillating terms conventionally disregarded in the rotating wave approximation (RWA). The tunneling amplitude displays resonant peaks as a function of the modulation frequency; near the maxima it is exponentially larger than the RWA tunneling amplitude. 
\end{abstract}

\pacs{03.65.Xp, 42.50.Pq, 74.78.Na,  05.60.Gg}

\maketitle

\section{Introduction}

Many systems of current interest can be modeled by modulated nonlinear quantum oscillators. Examples range from  Josephson junction based
systems \cite{Vijay2009,*Mallet2009,*Wilson2010} to
optical cavity modes \cite{Walls2008}, electrons in a Penning trap \cite{Peil1999}, and opto- and nano-mechanical systems 
\cite{Kippenberg2008,*Brennecke2008,*Clerk2010a,*Purdy2010,*Chan2011}. The oscillator dynamics is often characterized by well-separated 
time scales: the reciprocal  eigenfrequency $\omega_0^{-1}$
and a much longer time related to the vibration decay and nonlinearity. A standard approach to the analysis of the dynamics is based
 on the rotating wave approximation (RWA), where one separates  slow variables, like the vibration amplitude and the slow part of the phase,
and disregards the effect of fast oscillating terms on their evolution. 

An important quantum effect in modulated systems is dynamical tunneling \cite{Davis1981}. It can be understood for a parametric oscillator, which is excited by modulation at frequency $\omega_F$ close to $2\omega_0$. Classically, a weakly nonlinear oscillator can have two states of  vibrations at frequency $\omega_F/2$, which have the same amplitudes and differ in phase by $\pi$ \cite{LL_Mechanics2004}. Quantum fluctuations  cause tunneling between these states \cite{Wielinga1993,Marthaler2006,*Marthaler2007a}. Similar tunneling, which should be distinguished from dissipative switching \cite{Drummond1980c,*Kinsler1991,Dykman1988a}, is known also for other types of  
vibration bistability \cite{Sazonov1976,*Dmitriev1986a,*Vogel1988,*Peano2004,*Serban2007}.

In this paper, we show that the tunneling rate  of a 
parametrically modulated oscillator can be exponentially increased
by processes caused by the fast oscillating terms $\propto\exp(\pm in\omega_Ft)$, ($n= 1, 2, \ldots$)  disregarded
in the RWA. This happens where the difference of the appropriate eigenvalues of the RWA Hamiltonian becomes close to $n\hbar\omega_F$. The level configuration is of
$\Lambda$-type.  The two lowest RWA levels are degenerate (disregarding tunneling), with the wave functions localized near the period-two vibrational states, whereas the upper-level state is delocalized, see Fig.~1. The three states are resonantly mixed by the fast-oscillating non-RWA terms. The associated breakdown of the RWA is a purely quantum effect with no classical counterpart. 

The tunneling enhancement we consider is somewhat reminiscent  of photon-assisted tunneling from a potential well,  which is now broadly used in quantum information processing
\cite{Lucero2008}. There photon absorption resonantly accelerates tunneling decay if the photon energy $\hbar\omega_F$ coincides with the intrawell level spacing, 
since the decay rate of the excited state largely exceeds that of the ground state.
In contrast to systems displaying photon-assisted tunneling, a parametric oscillator is bistable due to the modulation, which forms the very barrier for tunneling in phase space. This leads to a different physics and requires a different description. 

\begin{figure}
\includegraphics[width=74mm]{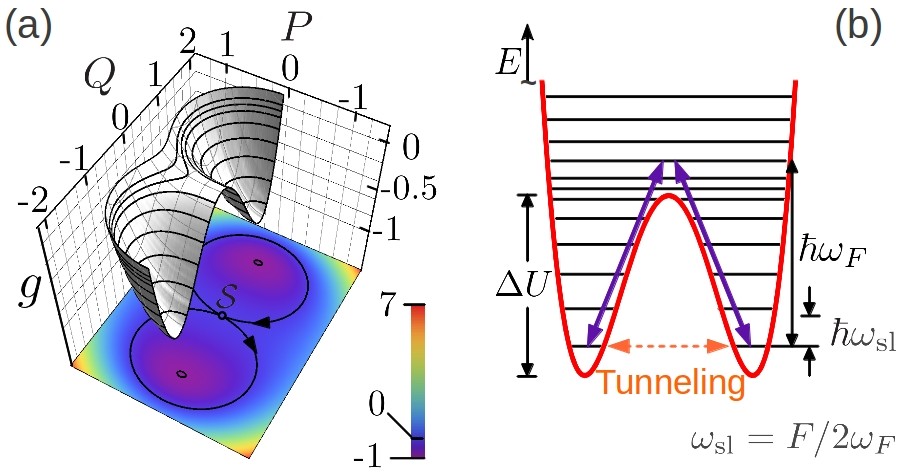}
\caption{(a) The dimensionless RWA Hamiltonian $g(Q,P)$, Eq.~(\protect\ref{eq:quasienergy}), for $\mu=0.95$. The minima of $g(Q,P)$ correspond to the parametrically excited vibrational states, in the presence of weak dissipation. (b) The cross-section $g(Q,P=0)$ with a sketch of RWA quasienergy levels; the arrows indicate resonant transitions due to the fast-oscillating corrections to the RWA. Also indicated are the typical energy scales. }
\label{fig1}
\end{figure}

We study moderately strong resonant modulation where the nonlinear part of the oscillator vibration energy remains small compared to the harmonic part. This makes the oscillator different from  modulated strongly nonlinear systems where much attention has attracted chaos-assisted \cite{Bohigas1993,*Hensinger2001c,*Steck2001} 
and nonlinear resonance-assisted tunneling \cite{Brodier2001,*Brodier2002,*Lock2010}.

For the rate of tunneling between the vibrational states to be small, the effective RWA tunneling barrier $\Delta U$ should largely exceed the RWA level spacing $\hbar\omega_{\rm sl}$; frequency $\omega_{\rm sl}$ characterizes the oscillator dynamics in the rotating frame, $\omega_{\rm sl}\ll\omega_F\approx 2\omega_0$.
The ratio $\Delta U/\hbar\omega_F$ can be arbitrary, it does not emerge in the RWA. We will be interested
in the case where $\Delta U\sim\hbar\omega_F$. In this case the effect 
of resonant admixture of the RWA states by the non-RWA interaction, is most pronounced,  see Fig.~1.

For $\omega_F\gg\omega_{\rm sl}$, the states resonantly mixed
by the non-RWA terms in the oscillator Hamiltonian overlap very weakly. We develop a method that allows us to calculate the relevant 
exponentially small matrix elements and to show that they are nevertheless sufficiently large to lead to exponential resonant enhancement of tunneling.

\section{RWA quasienergy states and their resonant mixing beyond the RWA}

The Hamiltonian of a parametrically modulated oscillator with   coordinate $q$, and 
momentum $p$ reads
\begin{equation}
H(t)=\frac{p^2}{2}+\frac{1}{2} q^2 [\omega_0^2 + F \cos(\omega_F
t)]+ \frac{\gamma}{4} q^4\,.  \label{ham}
\end{equation}
We assume that the  modulation  amplitude 
$F$ and the nonlinearity are comparatively  small, $F,\gamma\langle q^2\rangle\ll \omega_0^2$,  and the modulation frequency $\omega_F$
is close to  resonance, 
$|\omega_F- 2\omega_0|\ll \omega_0$; for concreteness we set $F,\gamma>0$.
Of primary interest is the range of the modulation parameters $F$ and $\omega_F$ where, 
 in the presence of weak damping, the oscillator has two almost sinusoidal stable classical vibrational states  with frequency $\omega_F/2$ \cite{LL_Mechanics2004}.

To study the oscillator dynamics, we  switch to the rotating frame using a standard
transformation 
$U(t)=\exp [-i\omega_F \hat{a}^\dagger \hat{a} t/2]$ and introduce dimensionless slow variables   $Q$ and $P$, $U^\dagger q U =C\left[
P\cos \omega_{\rm F} t/2-
Q\sin \omega_{\rm F} t/2\right]$ and $U^\dagger p U  =-(\omega_F/2)C\left[
P\sin \omega_{\rm F} t/2+
Q\cos \omega_{\rm F} t/2\right]$. Here $\hat{a}$ and $\hat{a}^\dagger$ are the ladder operators and $C  =(2 F/3 \gamma)^{1/2}$.
The Hamiltonian in the rotating frame $\tilde{H}=U^\dagger H U-i\hbar U^\dagger\dot{U}$ reads
\begin{eqnarray}
\label{eq:rot_frame_H}
\tilde{H}=(F^2/6\gamma)[g(Q,P)+ h(Q,P,t)].
\end{eqnarray}
The dimensionless
operator 
\begin{eqnarray}
\label{eq:quasienergy}
\hat{g}&=&\frac{1}{4}\left(Q^2+P^2\right)^2+\frac{1}{2}(1-\mu)P^2-\frac{1}{2}(1+\mu)Q^2\,
\end{eqnarray}
is independent of time \cite{Marthaler2006}. In contrast, the operator 
\[\hat{h}=h_1(Q,P) e^{- i\omega_F t}+h_2(Q,P) e^{-2i\omega_F t}+{\rm H.c.}\]
is
fast oscillating; $h_{1,2}$ are fourth
order polynomials in $Q$, $P$, they do not contain small 
parameters and are given explicitly in the Supplemental Material. Functions $g(Q,P)$ and $h(Q,P)$  are symmetric with respect to inversion $(Q,P)\to (-Q,-P)$ due to the periodicity of $H(t)$. They depend on a single dimensionless parameter $\mu$,
\begin{equation}
\label{eq:parameters}
\mu=\left[(\omega_F/2)-\omega_0\right]/\omega_{\rm sl}, \qquad \omega_{\rm sl}=F/2\omega_F.
\end{equation}
We disregard corrections $\sim\omega_{\rm sl}/\omega_F$.

The commutation relation for
the dimensionless coordinate $Q$ and momentum $P$  is 
\begin{equation}
\label{eq:lambda} 
[Q,P]=i\lambda,\qquad\lambda=3\gamma\hbar/(F\omega_F),
\end{equation}
where $\lambda$ is the dimensionless Planck constant. 
We assume that $\lambda\ll 1$. Then  quantum fluctuations are small on average.

From Eq.~(\ref{eq:rot_frame_H}), the Schr\"odinger equation  
in dimensionless time $\tau=t\omega_{\rm sl}$ is $i\lambda\partial_\tau\Psi= (\hat{g}+\hat{h})\Psi$. 
Since $\hat h$ is periodic in time, this equation has Floquet solutions
$\Psi_{\epsilon}(\tau+\tau_h)=\exp(-i \epsilon \tau_h/\lambda)\Psi_{\epsilon}(\tau)$. They
define the dimensionless quasienergies $\epsilon$ [$\tau_h=2\pi\omega_{\rm sl}/\omega_F\ll 1$].

In the RWA the fast oscillating term $\hat{h}$ is disregarded. Then operator $\tilde H$ becomes
time-independent. The dimensionless Hamiltonian
$g(Q,P)$, Eq.\ (\ref{eq:quasienergy}), is shown in Fig.~\ref{fig1}. It is not  a sum of the kinetic and potential energy. For $|\mu|<1$, $g(Q,P)$ has two symmetrically located minima, $g_{\rm min}=-(1+\mu)^2/4$, and a saddle point, $g_{\cal S}=0$.
 In the presence of weak dissipation, the minima correspond to the 
period-$2$ vibrational states in the laboratory frame, which have equal amplitude and opposite phase. The barrier 
height between the states
is $\Delta U=(F^2/6\gamma)(g_{\cal S}-g_{\rm min})$.

 The eigenvalues $g_m$ of $\hat{g}$ give dimensionless quasienergies $\epsilon_m$ in the RWA. For $\lambda\ll 1$ each well of $g(Q,P)$ 
in Fig.~\ref{fig1} contains many levels,
 $\propto 1/\lambda$. 
Because the wells are symmetric, the intrawell states are 
degenerate in the neglect of tunneling. With account taken of tunneling, the  eigenstates $\psi_n(Q)$ of $\hat g$ are even or odd in $Q$. The dimensionless RWA tunnel splitting $\delta g_{0}$ between the lowest-$g$ states was considered 
earlier \cite{Wielinga1993,Marthaler2006,*Marthaler2007a}. It is exponentially small, $|\log \delta g_{0} |\propto 1/\lambda$, and $\delta g_{0}$
oscillates with $\mu/\lambda$ \cite{Marthaler2007a}. 

The oscillating term $\hat h$ in the Hamiltonian (\ref{eq:rot_frame_H}) mixes RWA-eigenstates. For remote states the mixing is exponentially weak. However, it may become important where $\hbar\omega_F$ is close to the distance between the RWA levels, as it provides a new route for interwell transitions. Consider state $\psi_n$ above the barrier top with dimensionless quasienergy $g_n$ and the two lowest states $\psi^{(l)}_0$ and $\psi^{(r)}_0$ in the left and right wells of $g(Q,P)$ with quasienergy $g_0$ in the neglect of tunneling, see Fig.~\ref{fig1}(b). The dimensionless detuning between the interlevel distance and $\hbar\omega_F$ is $\Delta= \lambda^{-1}(g_n-g_0)-(\omega_F/\omega_{\rm sl})$. If $|\Delta|\ll 1$, transitions  $\psi^{(l,r)}_0\to \psi_n$ are resonant.  The matrix elements $\langle\psi_n|h_1|\psi^{(l)}_0\rangle$ and  $\langle\psi_n|h_1|\psi^{(r)}_0\rangle$ are equal for a symmetric  $\psi_n(Q)$ or have opposite signs for an antisymmetric $\psi_n(Q)$; we denote their absolute value by $h_{\rm res}$. To first order in $\hat h$, the amplitudes of resonantly coupled states  $\psi_n(Q),\psi^{(l,r)}_0(Q)$ oscillate at dimensionless frequencies
\begin{equation}
\label{eq:tunnel_frequencies}
\nu_{\mp}= [(\Delta^2+8\lambda^{-2} h_{\rm res}^2)^{1/2} \mp |\Delta|]/2.
\end{equation}
From Eq.~(\ref{eq:tunnel_frequencies}), $\nu_{\pm} \approx \sqrt{2}h_{\rm res}/\lambda$ for good resonance, $\lambda |\Delta|/h_{\rm res}\ll 1$. In the dispersive regime,  $\lambda|\Delta|/h_{\rm res}\gg 1$, interwell oscillations are characterized by frequency  $\nu_-\approx 2h_{\rm res}^2/\lambda^2\Delta$. As we show, in both cases $\nu_-$ can be exponentially larger than the dimensionless RWA tunneling frequency $\delta g_0/\lambda$, which was disregarded in Eq.~(\ref{eq:tunnel_frequencies}).

\section{Matrix elements for remote states}

The relevant matrix elements of $\hat h$ can be found using the WKB approximation, in the spirit of Ref.~\onlinecite{LL_QM81}. Interestingly, taking advantage of the conformal property of classical trajectories for the effective Hamiltonian $\hat g$, one can find both the exponent and the prefactor in the matrix elements. For the term $\propto h_1$ in $\hat h$ we write
\begin{eqnarray}
\label{eq:matrix_element_defined}
&& \langle\psi_ n|\hat{h}_{1}|\psi_0\rangle=2\mathrm{Re}\int_{0}^\infty dQ h^+(Q), \nonumber\\
&&h^+(Q)=\psi^+_n(Q) \hat h_{1}\psi_0(Q),
\end{eqnarray} 
where $\psi_0(Q)$ is one of the two tunnel-split lowest-$g$ states [the symmetric or antisymmetric combination of $\psi^{(l)}_0(Q)$ and  $\psi^{(r)}_0(Q)$ ]; $\psi_n(Q)$ has the same parity as $\psi_0$. Here we provide results for underbarrier states, $g_n< g_{\cal S}=0$; for $g_n>0$ the analysis is similar, see Supplemental Material.

Function $\psi^+_n(Q)$ is an eigenfunction of operator $\hat g$ such that 
$\mathrm{Re}[\psi_n^+(Q)]=\psi_n(Q)$ and, in the classically accessible region of the semiaxis $Q>0$, see Fig.~1,
\begin{eqnarray}
\label{eq:psi_n+}
 \psi^+_n(Q) \approx  c_n\left(\partial_P g_n\right)^{-1/2}\exp\left(i\lambda^{-1} S_n(Q) +i\pi/4\right).
\end{eqnarray}
Here, $S_n(Q) =\int_{a_R(g_n)}^QP(Q',g_n)dQ'$  is the mechanical action counted off from the right turning point $a_R(g_n)$
and $P(Q,g)$ is the classical momentum,
\begin{equation}
\label{eq:momentum}
P(Q,g)=\sqrt{\mu-1-Q^2+2\sqrt{g-g_{\rm min} - \mu+Q^2}}.
\end{equation}
In Eq.~(\ref{eq:psi_n+}) $\partial_Pg_n$ is $\partial_Pg$ calculated for $P=P(Q,g_n)$ and $c_n=[\tau_p^{(1)}(g_n)/2]^{-1/2}$, where $\tau_p^{(1)}(g)$ is the period of classical vibrations with quasienergy $g$.

\begin{figure}[h]
\includegraphics[width=72mm]{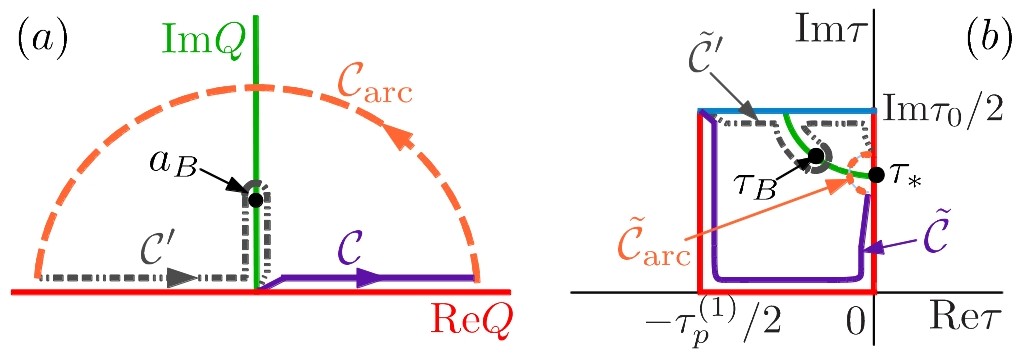}
\caption{(a) The contour of integration ${\cal C}$ for calculating the matrix element (\ref{eq:matrix_element_defined}) in the WKB approximation and the auxiliary integration contours for $\mu<0$; $a_B\equiv a_B(g_n)$ is the branching point of $P(Q,g_n)$,   $a_B(g)=i(g-g_{\rm min}-\mu)^{1/2}$, see Eq.~(\ref{eq:momentum}). (b) Mapping of the half-plane Im~$Q>0$ (with a branch cut) on the interior of a rectangle on the $\tau$-plane  for $\mu<0$ by function $Q(\tau;g)$ that describes the classical Hamiltonian trajectory with given $g$; $\tau_p^{(1)}$, $2\tau_0$, and $\tau_*$ are the real and imaginary periods and the pole of $Q(\tau;g)$, respectively. The solid ($\tilde{\cal C}$), dashed ($\tilde{\cal C}_{\rm arc}$), and dash-dotted ($\tilde{\cal C}'$) lines are the maps of the corresponding contours in (a). The arc in the upper right corner is the map of the imaginary axis of $Q$ from $a_B(g)$  to $i\infty$; $\tau_B$ is the time for reaching $a_B(g_n)$. }
\label{fig2} 
\end{figure}

We shift the integration path in Eq.~(\ref{eq:matrix_element_defined}) to the upper half-plane, contour ${\cal C}$ 
in Fig.~\ref{fig2}~(a). On this contour \cite{LL_QM81}
\begin{eqnarray}
\label{eq:h+WKB}
 h^+{}&& \approx [c_n c_0 h_{1}(Q,-P(Q,g_0))/2\sqrt{\partial_P g_n \partial_P g_0}]\nonumber\\
&&\times \exp\left\{i\left[S_n(Q)-S_0(Q)\right]/\lambda\right\}.
\end{eqnarray}
We then change from integration along ${\cal C}$ to integration along the semicircle  ${\cal C}_{\rm arc}$ at $|Q|\to \infty$, Im~$Q>0$, and contour ${\cal C'}$ that for $\mu < 0$ goes above the real axis from $-\infty$ to $Q=+0$ around the branch cut on the imaginary axis, see Fig.~\ref{fig2}~(a). We use that the classical trajectories $Q(\tau;g)$  for the Hamiltonian $g(Q,P)$ are expressed in terms of  the Jacobi elliptic functions \cite{Marthaler2006}. For each $g$, this expression provides conformal mapping of the half-plane Im~$Q >0$ (with a branch cut) onto a $g$-dependent region on the plane of complex time $\tau$. 
We define $\tau(Q,g)$ as the duration of classical motion from the turning point $a_R(g)$ to $Q$.
Then, for 
$\mu<0$ the region on the $\tau$-plane that corresponds to the half-plane Im~$Q>0$ is the interior of a rectangle shown in Fig.~\ref{fig2}~(b).

Using that $\tau(Q,g)=\partial S/\partial g$, we write the exponent in Eq.~(\ref{eq:h+WKB}) as
\begin{equation}
\label{eq:sn-s0}
\frac{i}{\lambda}[ S_n(Q)-S_0(Q)]=\frac{i}{\lambda}\int_{g_0}^{g_n}\!\!dg\tau(Q,g)\,.
\end{equation}
As seen in Fig.~\ref{fig2}~(b), for any $Q$ on contour   ${\cal C}_{\rm arc}$ and any $Q'$ on contour ${\cal C'}$,  Im~$\tau(Q,g) <$~Im~$\tau(Q',g)$.  Therefore the integral along  ${\cal C'}$ can be disregarded.

On contour ${\cal C}_{\rm arc}$  $\tau(Q,g)$ is given by the position of the pole  $\tau_*(g)$ of function $Q(\tau;g)$ \cite{Marthaler2006}, whereas from  Eqs.~(\ref{eq:quasienergy}) and (\ref{eq:momentum}) and the expression for  $h_1\bigl(Q,-P(Q,g_0)\bigr)$ (see Supplemental Material) the prefactor in $h^+$ is $\propto 1/Q$. Then
from Eq.~(\ref{eq:matrix_element_defined})
%
%
%
\begin{equation}
\langle \psi_n|\hat{h}_{1}|\psi_0\rangle\approx\frac{\pi}{3}c_n c_0 \exp\left[-\lambda^{-1}\int_{g_0}^{g_n} dg \;{\rm Im}~ \tau_*(g)\right].
\label{eq:matrix_element_explicit}
\end{equation}

\begin{figure}[h]
\includegraphics[width=82mm]{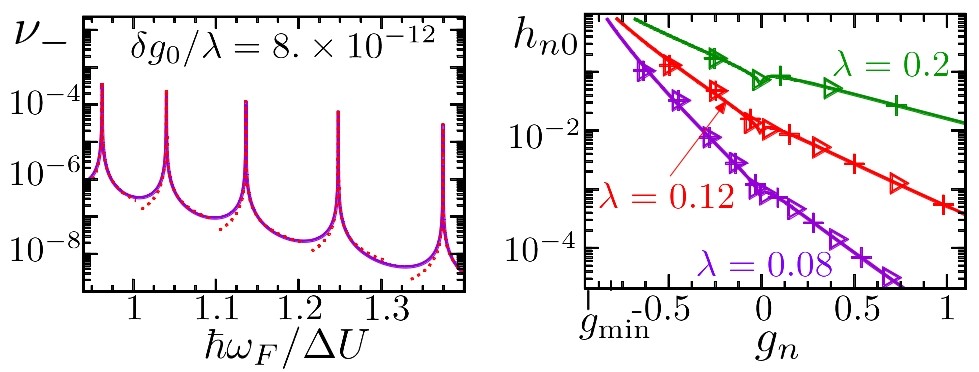}
\caption{Left panel: the scaled tunnel splitting $\nu_-=(\epsilon_1-\epsilon_0)/\lambda$ for the quasienergy states that maximally overlap with the lowest-$g$ states
 $(\psi^{(l)}_0\pm\psi^{(r)}_0)/\sqrt{2}$ for $\mu=0.95$ and $\lambda = 0.08$. 
The quasienergies $\epsilon_{0,1}$ are obtained numerically from the Schr\"odinger equation with the full Hamiltonian $H(t)$. The dotted lines show a comparison of the peak shapes with Eq.~(\protect\ref{eq:matrix_element_explicit}) for renormalized $g_n,h_{\rm res}$; the resonating $g_n$ are near the barrier top in Fig.~\ref{fig1}. Right panel: A comparison of Eq.~(\protect\ref{eq:matrix_element_explicit}) for $h_{n0}=\langle\psi_n|\hat h_1|\psi_0\rangle$ calculated as a continuous function of $g$ (the solid lines) with numerical calculations; the triangles and crosses refer to symmetric and anti-symmetric states $\psi_n$. A narrow vicinity of $g_n=0$ where expression (\protect\ref{eq:matrix_element_explicit}) logarithmically goes to zero is not shown.}
\label{fig3} 
\end{figure}

Equation  (\ref{eq:matrix_element_explicit}) gives the matrix elements of the fast-oscillating field $h_1$ in the explicit form, including both the exponent and the prefactor. An excellent agreement of Eq.~(\ref{eq:matrix_element_explicit}) with numerical calculations is seen in Fig.~\ref{fig3}. The matrix elements of $h_2$ are exponentially smaller than those of $h_1$ and can be disregarded. 

Expression (\ref{eq:matrix_element_explicit}) determines the matrix element $h_{\rm res}=|\langle\psi_n|\hat h_1|\psi_0\rangle/\sqrt{2}$ in Eq.~(\ref{eq:tunnel_frequencies}) for the interwell oscillation frequency. One should compare the exponent in Eq.~(\ref{eq:matrix_element_explicit}) with the RWA tunneling exponent $\log\delta g_0$. The latter is determined by action $S_0(-a_R(g_0))$ for moving  under the barrier from one well of $g(Q,P)$ to the other \cite{Marthaler2006}. By symmetry, it is given by twice the real part of Eq.~(\ref{eq:sn-s0}) for $g_n=0$ and $Q=+0$. From Fig.~\ref{fig2}~(b), Im~$\tau(0,g)={\rm Im}~\tau_0(g)/2 > {\rm Im}~\tau_*(g)$ for $g<0$. Therefore for $g_n<0$ not only $|\langle\psi_n|\hat h_1|\psi_0\rangle|$, but also $|\langle\psi_n|\hat h_1|\psi_0\rangle|^2$ are exponentially larger than $\delta g_0$. As shown in Supplemental Material, this relation holds also for $\mu>0$. 

\section{Discussion of results}

Of utmost interest for observing non-RWA tunneling is 
resonance of $\hbar\omega_F$ with states close to the barrier top in Fig.~\ref{fig1}. This is because the matrix elements (\ref{eq:matrix_element_explicit}) fall down exponentially with increasing $g_n$, whereas the condition of resonance with only one state $n$ requires that $|g_n-g_{n\pm 1}|\gg h_{\rm res}$; this condition is violated deep inside the wells of $g(Q,P)$, because tunnel splitting quickly becomes smaller than $h_{\rm res}$ with decreasing $g_n$. 

For $\hbar\omega_F\sim \Delta U$ one should take into account transitions $\psi^{(l,r)}_0\to \psi_n$ via intermediate states, which appear in higher-order in $\hat h$. From Eq.~(\ref{eq:matrix_element_explicit}), if the intermediate states are arranged in the order of increasing quasienergies, the resulting transition matrix elements have the same exponent as $h_{\rm res}$. Thus the corresponding virtual processes just renormalize the prefactor in the resonant tunnel splitting  compared to Eq.~(\ref{eq:matrix_element_explicit}). Figure~\ref{fig3} shows extremely sharp resonant peaks of the splitting of quasienergy levels for $\hbar\omega_F$ resonating with the RWA interlevel distance $(F^2/6\gamma)(g_n-g_0)$; this distance is also renormalized in higher orders in $\hat h$ and was used as an adjustable parameter.

Even moderately weak relaxation modifies the interwell transitions if the oscillator decay rate $\Gamma$ exceeds the tunneling frequency $\omega_{\rm sl}\nu_-$. We will consider the resonant case, where the dimensionless decay rate $\kappa=\Gamma/\omega_{\rm sl} \ll 1$ but $\kappa\gg \Delta$. If $\kappa\gg h_{\rm res}/\lambda$, resonant transitions $\psi^{(l,r)}_0 \to \psi_n$ occur at dimensionless rate $\sim h_{\rm res}^2/\lambda^2\kappa$. From the resonantly excited  state $\psi_n$ the system drifts down in quasienergy and approaches the states $\psi^{(l)}_0$ and $\psi^{(r)}_0$ with equal probabilities. As a result, instead of tunneling the system incoherently switches between the wells with rate $\sim h_{\rm res}^2/\lambda^2\kappa$, see Supplemental Material. 

Relaxation leads to interwell switching on its own via the mechanism of quantum activation \cite{Dykman1988a,Marthaler2006}. The dimensionless switching rate is $\nu_{QA}\sim \kappa\exp(-R_A/\lambda)$. From the explicit form of the activation exponent $R_A$ \cite{Marthaler2006} and Eq.~(\ref{eq:matrix_element_explicit}) it follows that, for $T=0$, the rate of $\hat h$-field induced switching is exponentially higher than the quantum activation rate for $\mu \gtrsim -0.35$, if the resonant quasienergy level is near the barrier top, $|g_n| \lesssim \lambda$. For $T$ exceeding a small $\mu$- and $\kappa$-dependent threshold value (still $T\ll \hbar\omega_F/2k_B$),  $R_A$ becomes smaller than the leading-order term in $2\lambda |\log h_{\rm res}|$. The difference between these quantities quickly falls down with increasing $\mu$, and for realistically not too small $\lambda$ and small quantum activation prefactor $ \kappa$, the $\hat h$-field induced switching may still dominate at resonance.

In conclusion, we have found a new mechanism of transitions between coexisting vibrational states of a parametric oscillator. The transitions correspond to resonant tunneling in a $\Lambda$-type configuration of quasienergy states and come from the terms, which are conventionally disregarded in the RWA. Using the conformal mapping technique, we show that the transition amplitude is exponentially larger than the RWA tunneling amplitude. It displays sharp resonant peaks as a function of the modulation frequency. These peaks should make it possible to observe the effect in experiments on oscillators with a high quality factor, including the currently studied Josephson junction based oscillators \cite{Vijay2009}.

The research of VP and MID was supported in part by the NSF, grant EMT/QIS 082985

\appendix

\section{Classical trajectories and conformal mapping}

The classical solutions $Q(\tau,g)$ for $g<0$ in terms of the Jacobi elliptic functions are given in Ref. \onlinecite{Marthaler2006}, except that here we count time off from the larger- rather than
the smaller-$Q$ turning point. In the main text we analyzed 
the conformal mapping of the half-plane Im~$Q >0$ and the integration contour in Fig.~2~(a) onto the $\tau$-plane for the case $\mu<0$.
The mapping changes for $g<-(1-\mu)^2/4$, in which case 
the momentum branching point $a_B(g)=i(g-g_{\rm min}-\mu)^{1/2}$ is on the real $Q$-axis. For the ground state, $g=g_0\approx g_{\min}$, this happens for $\mu>0$. On the other hand,  for the resonant quasienergy $g_n\approx g_{\min}+\hbar\omega_F/E_{\rm sl}$ close to the barrier top, $a_B(g_n)$ is imaginary for all $|\mu|\leq 1$. The mapping depends also on the sign of the parameter $m_J$ of the Jacobi elliptic functions, which itself depends on $\mu$ and $g$.

The mapping for $g<0$ and $\mu >0$ can be found in the same way as for $\mu < 0$. To find the mapping for $g>0$ we use the expression for the Hamiltonian trajectory $Q(\tau;g)$, which can be obtained by analytically continuing the result for $g < 0$,
\begin{eqnarray}
\label{eq:csol}
 Q(\tau;g)&=&\frac{2^{3/2}ig^{1/2}\dn \left(2(1+i)g^{1/4}\tau\right)}{\kappa_+-\kappa_- 
\cn\left(2(1+i)g^{1/4}\tau\right)}\,,
\end{eqnarray}
where $\kappa_\pm=\left(1+\mu\pm2ig^{1/2}\right)^{1/2}$ and
\begin{equation}
\label{mgg0}
 m_J(g) =1/2+i(1-\mu^2-4g)/8\sqrt{g}.
\end{equation}
From Eq.~(\ref{eq:csol}), the smallest in the absolute value position of the pole of $Q(\tau,g)$ in the upper half-plane of $\tau$ is 
\begin{equation}
\tau_*(g)=i\left| 2^{-3/2}g^{-1/4}\cn^{-1}(\kappa_+/\kappa_-)\right|\,.
\end{equation}

\begin{figure}
\begin{center}
\includegraphics[width=85mm]{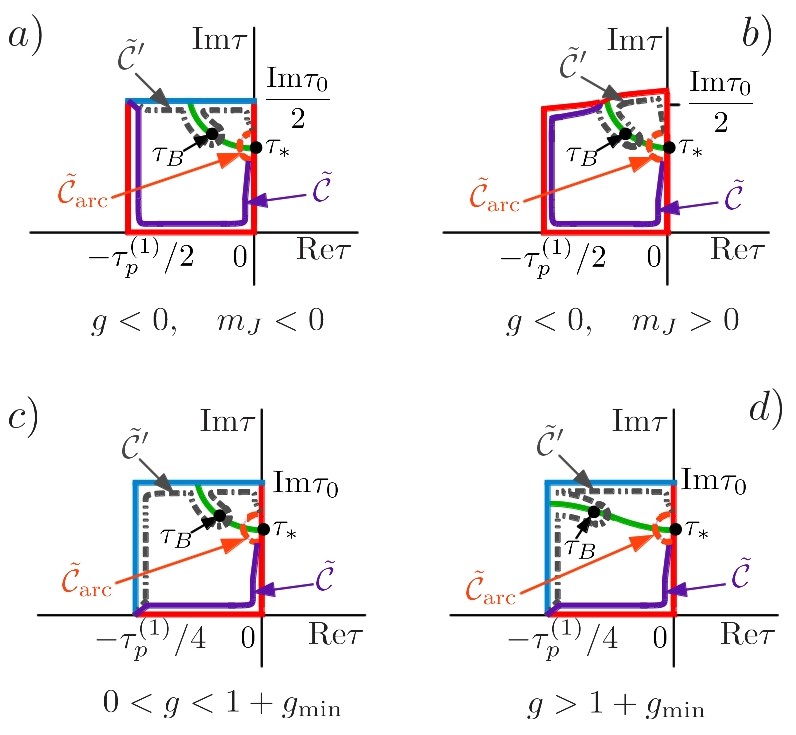}
\end{center}
\caption{\label{fig1s}  Conformal mapping of the half-plane Im~$Q>0$  (with the branch cut) on the 
$\tau$-plane using the classical Hamiltonian trajectories $Q(\tau;g)$ for underbarrier ($g<0$) and overbarrier ($g>0$) trajectories; in (a) and (b) $\mu <0$ and $\mu>0$, respectively; in (c) $0<g<1+g_{\rm min}$, and in (d)  $g>1+g_{\rm min}$. Parameters
$\tau_p^{(1)}$ and $2\tau_0$ are the real and complex periods of $Q(\tau;g)$, and 
$\tau_*$ is the smallest in the absolute value pole of $Q(\tau;g)$. The solid line $\tilde{\cal C}$, the dashed line $\tilde{\cal C}_{\rm arc}$, 
and the dash-dotted line $\tilde{\cal C}'$ show the mapping of the corresponding integration contours, see  Fig.~2~(a) 
of the main text. 
The green line
is the map of the imaginary $Q$-axis from the branching point $a_B(g)$ to $i\infty$; $\tau_B$ is the time for reaching $a_B(g_n)$. The blue line corresponds to moving  from $Q=-0$ to $Q=+0$ along the imaginary $Q$-axis around the branching point $a_B(g)$ in (a) and around $a_B(g)$ and the turning point $P(Q;g)=0$ in (c) and (d)}
\end{figure}

Figure \ref{fig1s} shows  the mapping onto the $\tau$-plane of the
 integration contours ${\cal C}$, ${\cal C'}$ and 
${\cal C}_{\rm arc}$ on the $Q$-plane plotted in Fig.~2 of the main text. It follows directly from the explicit expressions for $Q(\tau;g)$. Panels (a)-(d) refer to the four different possible situations. In all cases except for the one in Fig.~\ref{fig1s}~(b), the half-plane Im~$Q>0$ , with the branch cut along the imaginary axis, is mapped onto the interior of a rectangle. In Fig.~\ref{fig1s}~(b), the upper part of the rectangle is replaced with a curve that corresponds to going between the branching points of $P(Q,g)$ on the real $Q$-axis. Panel (a) coincides with Fig.~2(b) of the main text and is given for completeness.

The integral along  ${\cal C'}$ can be disregarded when calculating the matrix elements
$\langle\psi_n|\hat h_1|\psi_0\rangle$, 
if the integrand on  ${\cal C'}$ is exponentially smaller than the integrand on ${\cal C}_{\rm arc}$,
\begin{equation}
\label{eq:ineq}
\int_{g_0}^{g_n}dg\,\mathrm{Im}~\tau(Q_{\rm arc},g) <\int_{g_0}^{g_n}dg\,\mathrm{Im}~\tau(Q',g)
\end{equation}
for any $Q_{\rm arc}$ on contour   ${\cal C}_{\rm arc}$ and any $Q'$ on contour ${\cal C'}$. 

From inspection of Fig.\ \ref{fig1s}~(a), (b)
we see that the inequality holds for $g_n<0$, since
 Im~$\tau(Q_{\rm arc},g) <$ Im~$\tau(Q',g)$ for all $Q_{\rm arc}$ and $Q'$. The latter inequality fails for $g>0$ in a certain range of $Q'$. However, we emphasize that Eq.~(\ref{eq:ineq}) is only a sufficient condition for disregarding the integral along ${\cal C}'$, and condition Im~$\tau(Q_{\rm arc},g) <$ Im~$\tau(Q',g)$ is only a sufficient condition for Eq.~(\ref{eq:ineq}) to hold. Clearly, the matrix element $\langle\psi_n|\hat h_1|\psi_0\rangle$ is given by the integral along ${\cal C}_{\rm arc}$ for small $g_n>0$, where the major contribution to the integrals (\ref{eq:ineq}) comes from the range $g<0$. The range of small $|g_n|$ is of utmost interest for observing the non-RWA switching, as explained in the main text.

We note that in Eq.~(10) of the main text we disregarded the change of $\psi_0(Q)$ for $Q>0$ due to the tail of the part of this wave function  localized in the $Q<0$-well of $g(Q,P)$. In other words, we approximated $\psi_0(Q)=[\psi^{(r)}_0(Q)\pm \psi^{(l)}_0(Q)]/\sqrt{2}$ by $\psi^{(r)}_0(Q)/\sqrt{2}$ for $Q>0$ . This is certainly justified in the case of interest where the matrix element $\langle \psi_n|\hat h_1|\psi_0\rangle$ is exponentially large compared to the matrix element of tunneling between the wells of $g(Q,P)$ for $g=g_0$.

Our numerical analysis shows that the major contribution to the matrix element  $\langle \psi_n|\hat h_1|\psi_0\rangle$ comes from integration along ${\cal C}_{\rm arc}$ even for large $g_n>0$. This is is a consequence of fast oscillations of the integrand on the contour ${\cal C}'$.

\section{The prefactor in the WKB matrix elements}

It follows from Eqs.~(10) and (11) of the main text and the ensuing analysis that, in the WKB approximation, the matrix elements of operators $\hat h_{1,2}$ are given by integrals over an arc on the $Q$-plane that goes from $0$ to $\pi$ for $|Q|\to \infty$,
\begin{eqnarray}
\label{eq:complete_integral}
&&\langle\psi_ n|\hat{h}_{j}|\psi_0\rangle \approx  c_n c_0\mathrm{Re}\int_{{\cal C}_{\rm arc}}\!\!\!\!dQ\,
 h_{j}(Q,-P(Q,g_0))\nonumber\\
&&\times\left(\partial_P g_n \partial_P g_0\right)^{-1/2}\exp\left[\frac{i}{\lambda}
\int_{g_0}^{g_n}\!\!dg\tau(Q,g)\right]\,
\end{eqnarray}
where $j=1,2$. The explicit form of $\hat h_{1,2}$ is
\begin{eqnarray}
\label{eq:h_operators}
    \hat{h}_1&=&(P-iQ)\left[(Q^2+P^2)/6 -\mu/4\right](P-iQ) \nonumber\\
&&+(Q^2+P^2)/2,\nonumber\\
     \hat{h_2}&=&(P-iQ)^4/24 + (P-iQ)^2/4.
\end{eqnarray}  
The  normalization constant $c_n$  in Eq.~(\ref{eq:complete_integral}) is $c_n=[\tau_p^{(1)}(g_n)/4]^{-1/2}$, for $g_n>0$.

On the integration contour in Eq.~(\ref{eq:complete_integral}), from Eqs.~(3) and (9) of the main text we have
$P(Q,g)\approx -i Q+i$, 
$\partial_P g\approx -2iQ^2$,  $\tau(Q,g)\approx\tau_*(g)$. Then from Eq.~(\ref{eq:h_operators}) $h_1\bigl(Q,-P(Q,g_0)\bigr)\approx 2 Q/3$ and $h_2\bigl(Q,-P(Q,g_0)\bigr)\approx -5/24$.  Substituting these expressions into Eq.~(\ref{eq:complete_integral}) gives Eq.~(12) of the main text for the matrix element of $\hat h_1$. In the case of $\hat h_2$ the integrand in Eq.~(\ref{eq:complete_integral}) is $\propto 1/Q^2$ for $|Q|\to \infty$, so that the integral (\ref{eq:complete_integral}) goes to zero. Therefore the matrix element $\langle\psi_n|\hat h_2|\psi_0\rangle$ is determined by the exponentially smaller contribution
that comes from integration along the path ${\cal C'}$.

\section{Effects of dissipation}

In the presence of dissipation, the dynamics of the oscillator in the rotating frame is described by the master equation for the density matrix $\rho$,
\begin{eqnarray}
\label{eq:QKE}
&&\partial_{\tau}\rho = i\lambda^{-1}\left[\rho, \hat g + \hat h\right] - \hat\kappa\rho,\qquad\hat\kappa\rho  = \kappa(\bar n + 1 )\nonumber\\
&&\times(a^{\dagger}a \rho - 2 a\rho a^{\dagger}  + \rho a^{\dagger}a) + \bar n (aa^{\dagger} \rho - 2 a^{\dagger} \rho a + \rho a a^{\dagger}).
\end{eqnarray}
Here, $\kappa=\Gamma/\omega_{\rm sl}$ is the scaled coefficient of viscous friction (the classical friction force is $-2\Gamma dq/dt$) and $\bar n = \left[\exp(\hbar\omega_F/2k_BT)-1\right]^{-1}$ is the oscillator Planck number; $a^{\dagger}$ and $a$ are the raising and lowering operators, $a=(2\lambda)^{-1/2}(Q+iP)$.


We will consider $\hat h$-induced switching between the period-two vibrational states in  the most interesting case of moderately small decay rate, $1\gg \kappa\gg h_{\rm res}$. For resonant transitions $\psi^{(l,r)}_0\to \psi_n$ the detuning $\Delta= \lambda^{-1}(g_n-g_0)-(\omega_F/\omega_{\rm sl})$ will be assumed small, $|\Delta|\ll \kappa$. 
If the system is initially prepared in one of the wells of $g(Q,P)$, over dimensionless time $\sim\kappa^{-1}$ the distribution $\rho$ becomes quasi-stationary, except for the well populations, which as we show vary on a much longer time.
%

For $\kappa\ll 1$ off-diagonal matrix elements of $\rho$ on the eigenfunctions of operator $\hat g$ are small. They can be disregarded except for the matrix elements $\rho^{(r,l)}_{n0}=\langle \psi_n|\rho|\psi^{(r,l)}_0\rangle$ for the resonant state $\psi_n$ and their conjugate (the other exception are matrix elements in a narrow range of $g$-values near the barrier top where the tunnel splitting is $\sim \lambda\kappa$). 
From Eq.~(\ref{eq:QKE}),  to the leading order in $h_{\rm res}/\kappa$, in the quasi-stationary regime
\begin{eqnarray}
\label{eq:off_diagonal}
\tilde\rho^{(r)}_{n0} \approx -i \frac{h_{\rm res}/\lambda}{\kappa(W_0 + W_n)} \left(\tilde\rho^{(rr)}_{00}-\tilde\rho_{nn}\right)
\end{eqnarray}
and similarly for $\tilde\rho^{(l)}_{n0}$. Here we use the interaction representation, $\tilde\rho = V^{\dagger} \rho V$ with $V\equiv V(\tau)=\exp(-i\hat g \tau/\lambda)$; $\tilde \rho_{nn}$ and $\tilde\rho^{(rr)}_{00}$ are the diagonal matrix elements of $\tilde \rho$ on the wave functions $\psi_n$ and $\psi^{(r)}_0$, respectively.

Parameters $W_0$ and $W_n$ are determined by the rates of dissipation-induced transitions from states $\psi^{(r)}_0$ and  $\psi_n$. From Eq.~(\ref{eq:QKE}),
\begin{eqnarray}
\label{eq:transition_rates}
&&W_{m}=\sum\nolimits_{m'}W_{mm'}, \nonumber\\ 
&&W_{mm'}=(\bar n + 1)|\langle\psi_{m'}|a|\psi_m\rangle|^2 + \bar n |\langle\psi_m|a|\psi_{m'}\rangle|^2.
\end{eqnarray}
Matrix elements $W_{mm'}$ were calculated before \cite{Marthaler2006}; they exponentially decay with increasing $|m-m'|$, as is also clear from Eq.~(12) of the main text. In Eq.~(\ref{eq:off_diagonal}) we neglected direct tunneling between the states $\psi^{(r)}_0$ and $\psi^{(l)}_0$.

The matrix elements (\ref{eq:off_diagonal}) determine influx into the population of the resonant state $\tilde\rho_{nn}$.  If initially the oscillator is in the right well of $g(Q,P)$, for example, then $\tilde\rho^{(rr)}_{00}\gg \tilde\rho^{(ll)}_{00}, \tilde \rho_{nn}$ for not too large $\tau$, and from Eq.(\ref{eq:QKE}) the influx  term is
\begin{eqnarray}
\label{eq:influx}
&&[\partial_{\tau}\tilde\rho_{nn}]_{\rm in} = -2\lambda^{-1} h_{\rm res} {\rm Im}~\tilde\rho^{(r)}_{n0}\approx 2\nu_h \tilde\rho^{(rr)}_{00}, \nonumber\\
&&\nu_h=\lambda^{-2}h_{\rm res}^2/[\kappa(W_0 + W_n)]. 
\end{eqnarray} 

Dissipation leads to transitions from the resonant state $\psi_n$ to neighboring states $\psi_m$, with rates $2\kappa W_{nm}$. They result in a current in the space of state populations. The current density can be found from the balance equation for the diagonal matrix elements $\tilde\rho_{mm}$ assuming that $\tilde\rho_{mm}$ is a smooth function of $m$, i.e., $\tilde\rho_{m+k,m+k}\approx \tilde \rho_{mm}+ k\partial_m\tilde\rho_{mm}$ for $|k|\ll m$. 

In the WKB approximation, the matrix elements $W_{mm'}$ for small $|m-m'|$ are expressed in terms of the Fourier components of the periodically oscillating in time momentum $P$ and coordinate $Q$ on a classical trajectory with Hamiltonian $g(Q,P)$ \cite{LL_QM81}. It is straightforward to show that, for a smooth $\tilde\rho_{mm}$,
\begin{equation}
\label{eq:current}
\sum_{m'}\left(W_{m'm}\tilde\rho_{m'm'}- W_{mm'}\tilde\rho_{mm}\right)\approx 
\lambda^{-1}\partial_m\left[I(m)\tilde\rho_{mm}\right],
\end{equation}
where $I(m)$ is the classical action for an orbit with $g(Q,P)=g_m$; from the Bohr quantization condition, $\partial_mI(m)=\lambda$. 

Equation $\partial_{\tau}\tilde\rho_{mm}=-(2\kappa/\lambda)\partial_m[I(m)\tilde\rho_{mm}]$ has a standard form of a continuity equation. It applies for $g_m$ not too close to the resonant quasienergy $g_n$, the barrier top, and the minima of $g(Q,P)$. Inside the wells, for $|g_m|\gg \lambda$ and for the tunneling splitting small compared to $\lambda\kappa$, subscript $m$ enumerates intrawell states. For such states,  in the time range $\kappa^{-1}\ll \tau \ll \kappa(\lambda/h_{\rm res})^2$ the probability current $I(m)\tilde\rho_{mm}$ is constant. Since the wells of $g$ are symmetric and there is no accumulation away from the minima of $g$, the current is equal to half the influx into the resonant state (\ref{eq:influx}). Since the current into the initially empty well gives the rate of interwell switching, this rate is equal to $\nu_h$.

We note that, even for zero temperature $T$, dissipation leads to population of excited eigenstates of $\hat g$ and to interwell transitions via quantum activation. For $T=0$ the quantum activation exponent is $R_A= 2{\rm Im}~\int _{g_{\min}}^0dg[\tau_0(g)-2\tau_*(g)]$  \cite{Marthaler2006}. Comparing this integral with the exponent of $\nu_h\propto h_{\rm res}^2$ we obtained the range of $\mu$ where switching via a resonant state is exponentially more likely than quantum activation.


%

\end{document}